\documentclass[10pt,prx,aps,amsfonts,floatfix,superscriptaddress,longbibliography,twocolumn,footinbib]{revtex4-1}
\usepackage[utf8]{inputenc}
\usepackage{amsmath}    
\usepackage{amssymb}
\usepackage{mathbbol}
\usepackage{graphicx}   
\usepackage{verbatim}   
\usepackage{subfigure}  
\usepackage{hyperref} 
\usepackage{scalerel}
\usepackage{cancel}
\usepackage{soul}
\usepackage[normalem]{ulem}
\usepackage[dvipsnames]{xcolor}
\usepackage{siunitx}
\usepackage{physics}
\usepackage{enumerate} 
\usepackage{empheq}

\usepackage{orcidlink}
\usepackage[draft,inline,nomargin]{fixme}
\fxsetup{theme=color}
\newcommand{\unam}{Instituto de Ciencias Nucleares, Universidad Nacional Aut\'onoma de M\'exico, Apdo. Postal 70-543, C.P. 04510  Cd. Mx., Mexico}

\newcommand{\icn}{Instituto de Ciencias Nucleares, \unam}
\newcommand{\pcsadd}{Center for Theoretical Physics of Complex Systems, Institute for Basic Science(IBS), Daejeon 34126, Republic of Korea}
\FXRegisterAuthor{mg}{amg}{\color{blue}MG}
\FXRegisterAuthor{mp}{prado}{\color{red}MiguelP}

\begin{document}

\title{Participation Ratio as a Quantum Probe of Hierarchical Stickiness}

\author{Ariel A. Galindo Duque,\orcidlink{0000-0001-7996-3739}}
\email{ariel.galindo@ciencias.unam.mx}
\affiliation{\icn}

\author{Miguel A. Prado Reynoso,\orcidlink{0009-0000-8983-1158}} 
\email{miguelangel.prado@urjc.es}
\affiliation{\icn}
\affiliation{Nonlinear Dynamics, Chaos and Complex Systems Group, Departamento de F\'{i}sica,  Universidad Rey Juan Carlos, Tulip\'{a}n s/n, 28933 M\'{o}stoles, Madrid, Spain}

\author{Miguel Gonzalez\,\orcidlink{0009-0004-0112-5988}}
\email{miguel.gonzalez@correo.nucleares.unam.mx}
\affiliation{\icn}
\affiliation{\pcsadd}

\author{Jorge G. Hirsch,\orcidlink{0000-0002-2170-9903}}
\email{hirsch@nucleares.unam.mx}
\affiliation{\icn}

\begin{abstract}
We investigate how quantum localization encodes the hierarchical stickiness that governs transport in mixed classical phase spaces. Using the periodically driven kicked top, we show that the participation ratio (PR) of coherent states in the Floquet eigenbasis quantitatively reflects the hierarchical organization of sticky regions identified classically through finite-time Lyapunov exponent (FTLE) statistics. To establish a quantitative quantum–classical correspondence, we introduce a Gaussian coarse graining of the FTLE matched to the intrinsic semiclassical resolution of coherent states. Both local correlations and global comparisons of probability distributions demonstrate that quantum and classical indicators agree optimally within a finite window of evolution times, where sticky structures are most clearly resolved. Our results promote the participation ratio from a global measure of chaos to a sensitive probe of hierarchical transport and provide a practical framework for diagnosing anomalous localization in driven quantum systems.
\end{abstract}

\maketitle

\textbf{This study examines how hierarchical stickiness in mixed classical phase spaces is encoded in quantum localization. We show that the participation ratio of coherent states in the Floquet eigenbasis quantitatively reflects the hierarchical organization of sticky regions within the chaotic sea, closely following the multimodal distribution of finite-time Lyapunov exponents. By matching classical and quantum resolutions, an optimal time window emerges in which the participation ratio and the coarse-grained finite-time Lyapunov exponent show maximal agreement. Complementary dynamical analysis further reveals distinct localization and spreading behaviors for coherent states initialized in different sticky regions. These results promote the participation ratio from a global measure of chaos to a sensitive probe of hierarchical transport.}

\section{Introduction}
\label{Intro}

Understanding how classical phase-space structures manifest themselves in quantum dynamics remains a central problem of quantum chaos. In systems with mixed phase space, remnants of invariant tori and partial transport barriers organize chaotic transport and generate long trapping times, a phenomenon commonly known as stickiness.
\cite{Contopoulos1971,Contopoulos2008, Contopoulos2010}. These sticky regions play a fundamental role in anomalous transport, leading to nonuniform exploration of the chaotic sea and the emergence of multiple characteristic dynamical behaviors.

Classically, stickiness \cite{Contopoulos1971,Contopoulos2008, Contopoulos2010,Dvorak1999,PRADOREYNOSO2022106358,REYNOSO2021110640,Livorati2018,Borin2023, Voglis2006} is efficiently characterized through finite-time Lyapunov exponents (FTLEs). Rather than forming a single chaotic component, FTLE distributions typically display multiple modes associated with nested layers of instability \cite{You2021, Stefanski2010, Drotos2021, Prasad1999, Manos2015, Prado2022, Storm2024,OttBook,Lapeyre2002, Sales2023, Prado2022b, Shadden2005}.
Such multimodality reflects the hierarchical organization of transport in phase space. Whether and how this hierarchy is encoded in quantum observables remains largely unexplored.

The participation ratio (PR) has become a standard observable for quantifying quantum-state delocalization and discriminating between regular and chaotic dynamics \cite{PhysRevE.93.022215,Quantum6.644.2022,PhysRevResearch.3.023214,PhysRevE.110.L062201}. More generally, the PR has been shown to reveal relevant classical and quantum structures, including fixed points, unstable periodic orbits, quantum scars, and symmetry-related features \cite{Ruebeck2017,Olibeira2018, Sieberer2019, ChavezPrado2025, PhysRevE.93.022215,PhysRevE.107.054213}. In particular, in mixed systems it has been successfully used to distinguish regular from chaotic eigenstates \cite{PhysRevE.93.022215,Olibeira2018,Sieberer2019,ChavezPrado2025,PhysRevE.107.054213}.

Here we demonstrate that the PR contains substantially finer information: it resolves the hierarchical organization of the chaotic sea itself and quantitatively reproduces the layered structure encoded in finite-time Lyapunov exponent statistics.

Periodically driven systems with a clear classical limit provide an ideal setting for addressing this question \cite{Moore1995,Mudute-Ndumbe2020}. Among them, the kicked top offers a paradigmatic framework in which coherent states furnish localized semiclassical probes while quantum evolution is governed by a well-defined Floquet operator \cite{PhysRevE.107.054213, quantumrecurrences2024, Pseudoclassical2022,Bastidas2014, Chaudhury2009,Krithika2019}. Moreover, the kicked top constitutes a representative model of Hamiltonian systems with mixed phase space, where regular islands coexist with chaotic transport and sticky structures emerge naturally.

In this work, we analyze coherent states expanded in the Floquet eigenbasis and show that their participation ratio reproduces the multimodal structure of the classical FTLE distribution. To make the correspondence quantitative, we introduce a Gaussian coarse graining of the FTLE matched to the intrinsic semiclassical resolution of coherent states. Using both pointwise and distributional measures, we identify an optimal time window in which classical hierarchical structures are most faithfully encoded in quantum localization.

Our results promote the participation ratio from a global indicator of chaos to a sensitive probe of hierarchical transport and provide a practical diagnostic of anomalous localization in driven quantum systems. Although the present work focuses on this specific system, the mechanisms explored here, are expected to be relevant more broadly in mixed Hamiltonian systems exhibiting partial transport barriers.

The paper is organized as follows. In Sec.~II, we introduce the classical and quantum kicked top, define the effective Planck constant, and construct coherent states on the sphere. In Sec.~III, we introduce the participation ratio in the Floquet basis and discuss its role as a global indicator of the regular–chaotic transition. Section~IV analyzes classical stickiness through finite-time Lyapunov exponents and demonstrates how quantum stickiness emerges in participation-ratio distributions. Section~V examines the dependence of PR–FTLE correlations on the FTLE time window and on $\hbar_{\mathrm{eff}}$. Finally, Sec.~VI summarizes our conclusions and discusses possible extensions to higher-dimensional and many-body systems.

\section{Regular–Chaotic Transition in the kicked top model} 
\label{model}

\subsection{The kicked top model}

The kicked top is a paradigmatic model for exploring chaotic dynamics in both classical and quantum mechanics. The quantum system is described by an angular momentum vector $\hbar \hat{\mathbf{J}} = \hbar(\hat{J}_x, \hat{J}_y, \hat{J}_z)$, whose components satisfy the $SU(2)$ commutation relations  $[\hat{J}_i,\hat{J}_j]=i\varepsilon_{ijk}\hat{J}_k$. In the setup we study, the system's dynamics are governed by the time-dependent Hamiltonian
\begin{equation}
\hat{H}(t)= \hbar\alpha\hat{J}_z + \frac{\hbar k}{2J}\hat{J}_x^2\sum_{n=-\infty}^\infty \delta\left(t-n\right)
\label{Hq}
\end{equation}
where the first term describes free precession around the $z$-axis with frequency $\alpha$, while the second term represents periodic kicks applied along the $x$-axis. The period between successive kicks has been set to unity. Each kick induces an instantaneous rotation about the $x$-axis by an angle proportional to $J_x$, with the proportionality factor given by the dimensionless coupling constant $k$. 
By applying Floquet's theorem, we can describe the evolution of the system in terms of the one-period evolution operator
\begin{equation}
\hat{U}=\exp\left(-i\alpha\hat{J}_z\right)\exp\left(-\frac{ik}{2J}\hat{J}_x^2\right)
\label{FO}
\end{equation}
where we set $\hbar=1$. The squared angular momentum operator, $\hat{J}^2$, is a conserved quantity satisfying $[\hat{J}^2, \hat{H}(t)] = 0$ and $[\hat{J}^2, \hat{U}] = 0$,
with eigenvalues $J(J+1)$.
As a result, one can describe the dynamics at a fixed value of $J$. 

The $(2J + 1)$-dimensional Hilbert space is spanned by the simultaneous eigenstates of $\hat{J}^2$ and $\hat{J}_z$, the well-known Dicke base. The spectral decomposition of the operator (\ref{FO}), $\hat{U}|\phi_j\rangle=e^{i\phi_j}|\phi_j\rangle$, yields the $(2J\!+\!1)$ Floquet eigenstates $|\phi_j\rangle$ and their corresponding quasi-energies $\phi_j$.

The classical Hamiltonian can be obtained from the quantum Hamiltonian (\ref{Hq}) by taking the expectation value with respect to spin-coherent \cite{PhysRevB.104.104409} states, as explained in Appendix \ref{ap1}:
\begin{equation}
\begin{split}
H \!\!=\! \alpha J_z \!+ \frac{k}{2} J_x^2 \sum_{n=-\infty}^\infty \!\!\!\!\delta\!\left(t - n\right)
\end{split}
\label{Hcl}
\end{equation}
The parameter $k$ controls the strength of the kicks and acts as the nonlinearity parameter of the system. 

By integrating Hamilton's equations of (\ref{Hcl}) over one period, we can obtain the stroboscopic evolution of the system, described by the iterated map

\begin{equation}
\begin{split}
J_{n\!+\!1}^x & \!\! = \! 
    \cos\alpha J_n^x  - \sin\alpha \left[J_n^y  \cos\left(kJ_n^x\right) \!+\! J_n^z  \sin\left(kJ_n^x\right)\right] \\
J_{n\!+\!1}^y &\!\!=\! 
    \sin\alpha J_n^x  + \cos \alpha \left[J_n^y \cos \left(kJ_n^x\right) \!-\! J_n^z \sin\left(kJ_n^x\right)\right] \\
J_{n\!+\!1}^z & \!\!=\!
    J_n^y \sin \left(kJ_n^x\right) + J_n^z \cos\left(kJ_n^x\right)
\end{split}
\end{equation}

\begin{figure*}[t]
    \centering
    \includegraphics[width=1.0\linewidth]{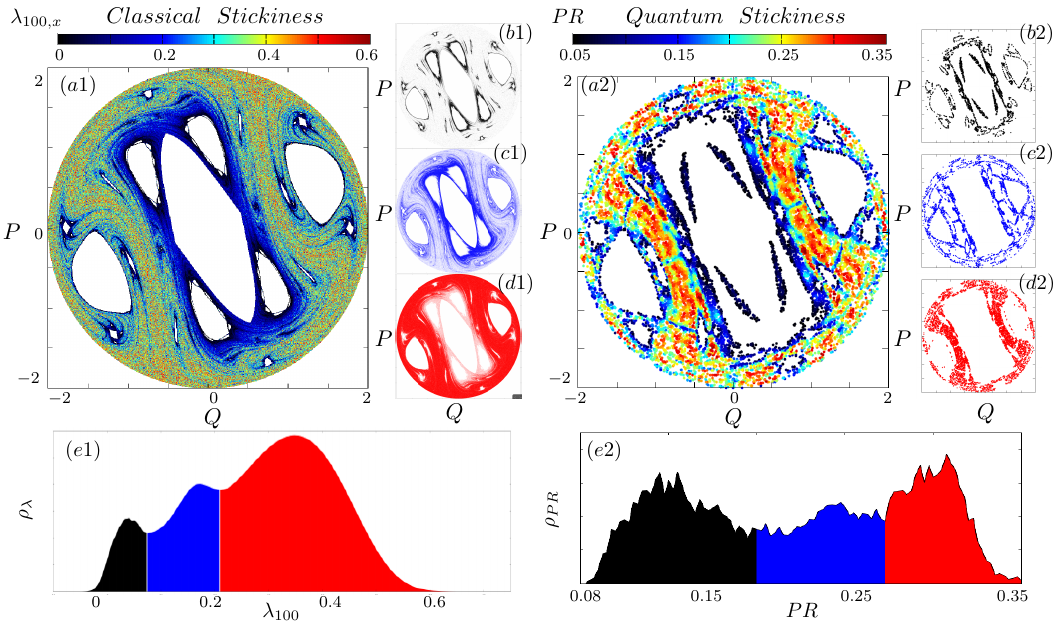}
    \caption{Hierarchical stickiness in the kicked top for $k=3$.
(a1) Finite-time Lyapunov exponent (FTLE) distribution over the chaotic sea in classical phase space.
(b1–d1) Phase space partitioned according to intervals of FTLE values. 
(e1) Multimodal histogram of the FTLE density.
(a2) Distribution of the participation ratio (PR) of coherent states
in phase space for $J=500$. (b2–d2) Phase space partitioned according to intervals of PR values.
(e2) Multimodal histogram of the PR density.
}
    \label{fig01}
\end{figure*}

\subsection{The participation ratio as a quantum signal of chaos}

The participation ratio (PR) is a standard measure of quantum-state localization in a chosen basis and is directly connected to the second-order Rényi entropy \cite{784q-xqmm, PhysRevE.109.034202,WangRobnik2021}. 

For a pure state $|\psi\rangle$ expanded in an orthonormal basis $\{|\phi_j\rangle\}_{j=1}^N$ of dimension $N$, we define the \textit{normalized participation ratio} as
\begin{equation}
PR=\frac{1}{N}\left(\sum_{j=1}^N |\langle \phi_j|\psi\rangle|^4\right)^{-1},
\end{equation}
which takes values in the interval $PR\in[1/N,1]$, corresponding to complete localization on a single basis state and uniform delocalization over the full Hilbert space, respectively. 

We evaluate this quantity for spin coherent states [Eq.~(\ref{spin_c})], expressed in the Floquet eigenbasis of the kicked-top evolution operator [Eq.~(\ref{FO})]. By construction, the normalized participation ratio probes the local degree of delocalization of semiclassical states in phase space, providing a direct quantum signature of the underlying classical dynamics. 

Its use as a quantum indicator of the regular–chaotic transition is by now well established and has been extensively explored in a wide variety of quantum systems (see, e.g., Refs.~\cite{Olibeira2018, Sieberer2019, ChavezPrado2025, PhysRevE.93.022215}). In these studies, the PR has been shown to reliably distinguish between regular and chaotic regimes through systematic changes in the degree of localization of quantum states, as well as through its sensitivity to fixed points and unstable periodic orbits (UPOs) \cite{Pilatowsky2020,Pilatowsky2021,Villasenor2021}. 

In Appendix \ref{ap2} a detailed analysis of the regular-chaotic transition is presented. This transition is consistently captured by three complementary indicators:
the classical chaotic area fraction $\mu_k$, the quantum quasienergy level-spacing statistic $r$, and the phase space averaged participation ratio $\langle PR\rangle$. 
The latter captures the same transition through the localization properties of coherent states, without relying on spectral statistics.

\section{Classical and Quantum Manifestations of Hierarchical Stickiness}
In the present work we explore
the phenomenon of stickiness within the mixed regime, concentrating on the representative case of a kicking strength $k=3.0$. We fix $\alpha=0.84$.

Mixed phase spaces exhibit a hierarchical organization in which chaotic trajectories may remain trapped near remnants of invariant tori for extended times. This stickiness leads to anomalous transport and is reflected in the statistics of finite-time Lyapunov exponents (FTLEs). We show that the same hierarchical structure is resolved by the participation ratio (PR) of coherent states in the Floquet eigenbasis.

\subsection{
Classical hierarchy from finite-time Lyapunov exponents} 
\label{Cstick}
For each initial condition, we compute the finite-time Lyapunov exponent (FTLE), defined in Appendix \ref{ap3},
over different evolution times. In a uniformly hyperbolic system, the FTLE distribution is unimodal. In contrast, mixed systems exhibit a multimodal structure: distinct peaks correspond to regions with different instability levels.

Figure \ref{fig01}(a1) shows the phase-space distribution of the FTLE. The chaotic sea does not form a homogeneous component; instead, it is organized in layers surrounding regular islands. These layers are associated with progressively smaller Lyapunov exponents, indicating increasingly long trapping times. They are shown individually in Figs \ref{fig01}(b1, c1, d1).

The corresponding FTLE distribution [Fig. \ref{fig01}(e1)] displays multiple peaks. The largest exponent corresponds to strongly chaotic trajectories, while secondary peaks arise from sticky regions near partial transport barriers. This multimodality provides a quantitative fingerprint of hierarchical stickiness.

\subsection{Quantum signatures of stickiness in the Participation Ratio}
\label{QS}

To probe the quantum dynamics locally in phase space, we expand coherent states in the Floquet eigenbasis and compute their participation ratio.

Figure \ref{fig01}(a2) shows the PR mapped over phase space. Regular islands
are not colored. Within the chaotic sea the PR is not uniform: it exhibits structured variations mirroring the classical hierarchy. The individual layers are shown in Figs. \ref{fig01}(b2, c2, d2).

The PR distribution [Fig. \ref{fig01}(e2)] reveals multiple peaks analogous to those of the FTLE distribution. Regions corresponding to smaller classical Lyapunov exponents are associated with reduced quantum delocalization. Thus, the PR resolves not only the regular–chaotic transition but also the layered structure inside the chaotic component.

This correspondence indicates that quantum localization encodes classical transport barriers beyond the global distinction between order and chaos.


\subsection{Dynamics across sticky regions}

\begin{figure}[t]
    \centering
    \includegraphics[width=1.0\linewidth]{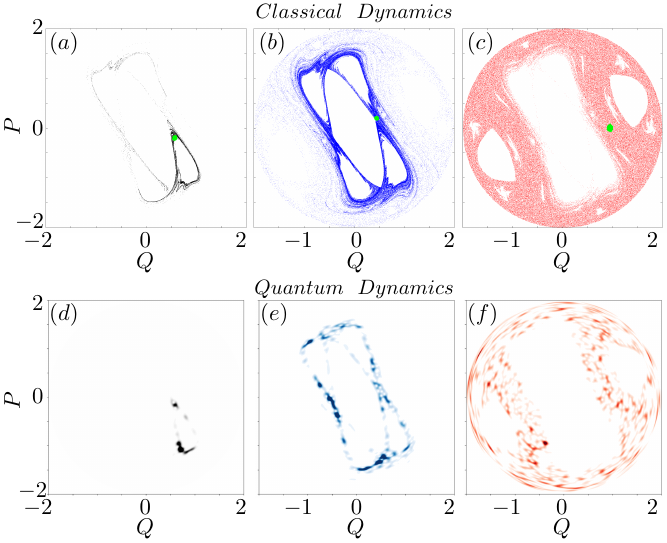}
    \caption{Dynamical evolution of coherent states in representative sticky regions of the chaotic sea for $k=3$. $(a)–(c)$ Classical phase-space evolution of a localized initial density after $100$ iterations. The green distribution denotes the initial Gaussian density on the unit sphere, chosen to mimic the phase-space localization of a coherent state. The three initial conditions correspond to regions with increasing instability: $(a)$ a strongly sticky region close to regular islands, characterized by small FTLE and PR values; $(b)$ an intermediate sticky region; and $(c)$ a more chaotic region with larger FTLE and PR values. The colored distributions show the evolved density after $100$ iterations. $(d)–(f)$ Husimi distribution of the corresponding coherent states evolved for the same time. Sticky regions exhibit comparatively more localized dynamics over the evolution time considered, while states initialized in strongly chaotic component exhibit enhanced spreading over phase space.}
    \label{figdyn}
\end{figure}

While the PR provides a static quantum signature of hierarchical stickiness, it is also instructive to examine how coherent states dynamically evolve when initialized in representative regions of the chaotic sea. To this end, we select three initial conditions associated with different levels of instability, identified through the FTLE and PR hierarchy discussed above.

Fig. \ref{figdyn} illustrates both the classical and quantum evolution of localized states after a fixed evolution time of $t=100$. The selected initial conditions correspond to: (a) a strongly sticky region located close to regular islands, characterized by comparatively small FTLE and PR values; (b) an intermediate sticky layer; and (c) a more strongly chaotic region associated with larger FTLE and PR values.

In the classical description [Fig.~\ref{figdyn}(a)–(c)], the initial localized density (green), chosen to reproduce the phase-space localization of a coherent state, evolves under the classical dynamics for 100 iterations. At the evolution time considered ($t=100$), the degree of spreading strongly depends on the underlying transport properties. Densities initialized in sticky regions [Figs.~\ref{figdyn}(a) and \ref{figdyn}(b)] remain temporarily confined near partial transport barriers, whereas those placed in more chaotic regions [Fig.~\ref{figdyn}(c)] spread more rapidly throughout the accessible phase space. At longer times, however, all three cases eventually explore the full chaotic component.

The corresponding quantum evolution is shown in Fig. \ref{figdyn}(d)–(f), where we plot the Husimi distribution of the evolved coherent states at the same evolution time. Coherent states initialized in highly sticky regions retain more localized structures over the evolution time considered, while states placed in more chaotic regions undergo stronger delocalization and develop broader interference patterns.

These results provide a dynamical interpretation of the PR hierarchy: the evolution of coherent states differs across the sticky regions identified through the FTLE and PR hierarchy, with smaller FTLE and PR values associated with more localized dynamics and larger values with more extended spreading over the chaotic sea.

The persistence of classical–quantum correspondence over comparatively long evolution times in sticky regions suggests that, in mixed phase-space systems, effective Ehrenfest-like times may depend on local finite-time instability rather than solely on the asymptotic Lyapunov exponent. A detailed analysis of this relation lies beyond the scope of the present work and will be explored in future studies.

\section{
Quantitative Quantum–Classical Correspondence}

The qualitative agreement between FTLE and PR phase-space maps suggests a deeper quantitative relationship. We now compare classical instability and quantum delocalization using both local correlations and global distributional measures.

\subsection{Resolution matching and Gaussian coarse graining}

A direct comparison between FTLE and PR requires accounting for their different phase-space resolutions. Classical trajectories probe arbitrarily fine structures, whereas coherent states provide a finite semiclassical resolution set by $\hbar_{\mathrm{eff}}=1/J$.

To establish a consistent comparison, we introduce a Gaussian coarse graining of the FTLE field (GFTLE), as explained in Appendix D.
The width is chosen to match the intrinsic spread of coherent states. This procedure filters classical structures below quantum resolution while preserving the hierarchical organization.

\begin{figure}
    \centering
    \includegraphics[width=1.0\linewidth]{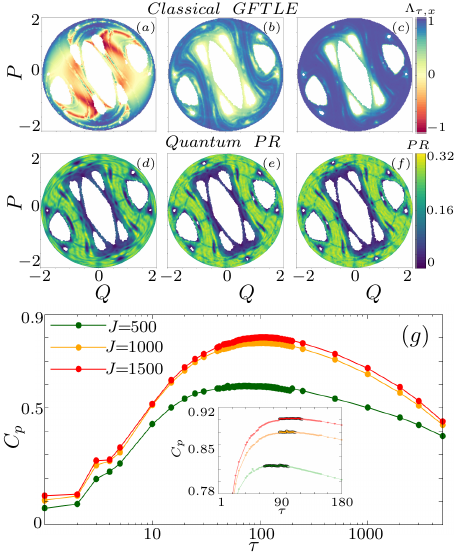}
    \caption{
    (a)-(c) GFTLE distribution 
    using a Gaussian smoothing width $\sigma^2=1/J$ for
    time windows $\tau=1, 100$, and $5000$.
    (d)-(f) PR distribution of coherent states
    for $J=500, 1000$, and $1500$.
    (g) Pearson correlation coefficient between the GFTLE and the PR.
    }
    \label{fig02}
\end{figure}

Figures \ref{fig02}(a, b, c) show
the coarse-grained GFTLE map for three  different times: $\tau=1, 100$ and $5000$, while Figs \ref{fig02}(d, e, f) display the PR distribution for $J=500, 1000$ and $1500$, illustrating the enhanced resolution of classical phase-space structures as $\hbar_{\mathrm{eff}}\sim J^{-1}$ decreases.

It can be seen that the GFTLE map for $\tau= 100$
closely resembles the PR landscape, including the layered structure within the chaotic sea.

\subsection{Local correlation analysis}

We quantify the pointwise correspondence using the Pearson correlation coefficient between the PR and the GFTLE.
Given two data sets
\[
A\!=\!\{A_i\}, \qquad
\!\!B\!=\!\{B_i\}\!,
\]
evaluated at identical phase-space points $(Q_i,P_i)$, we define the Pearson correlation coefficient as
\begin{equation}
C_{p}=\frac{\mathrm{Cov}(A,B)}{\sigma_A\sigma_B},
\end{equation}
where $\mathrm{Cov}(A,B)$ denotes the covariance between $A$ and $B$, and $\sigma_A$, $\sigma_B$ are their respective standard deviations. 

Figure \ref{fig02}(g) shows the correlation as a function of evolution time $\tau$. The agreement is not monotonic: correlations increase for intermediate times and decrease for both short and long times. Short times do not resolve transport barriers, while long times smear hierarchical distinctions. An optimal time window emerges in which classical and quantum indicators are maximally aligned.

The dependence on spin size 
$J$ [Fig.\ref{fig02}(g)] is consistent with the semiclassical origin of the correspondence. Increasing $J$ enhances resolution and sharpens correlations, consistent with the expectation that quantum localization increasingly reflects classical structures as $\hbar_{eff} \rightarrow 0$.

\subsection{Comparison of distributions}

\begin{figure}
    \centering
    \includegraphics[width=1.0\linewidth]{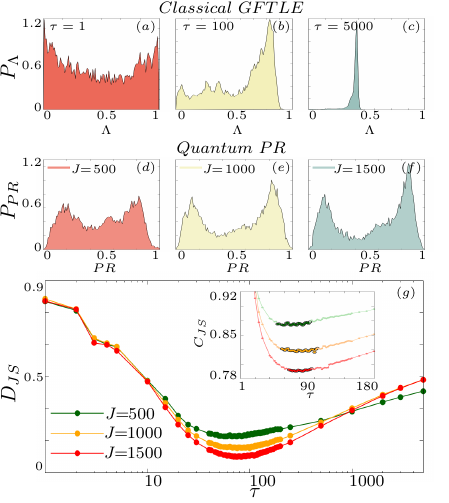}
    \caption {(a)-(c) GFTLE probability distributions
    for $\tau=1, 100$, and $5000$.
    (d)-(f) PR probability distributions for $J=500, 1000$, and $1500$. (g)
    Jensen--Shannon distance between the GFTLE and PR
    as a function of the time window $\tau$ (logarithmic scale). Inset: magnified view of the region around the minimum, in linear scale.}
    \label{fig03}
\end{figure}

In Figs.~\ref{fig03}(a)-(c), we plot the 
histogram of the density Gaussian-smoothed finite-time Lyapunov exponent (GFTLE) for three different time windows, $\tau=1, 100$, and $5000$, respectively. Figures~\ref{fig02}(d)--(f) display the quantum counterpart, the histogram of the PR density
for three different values of the spin magnitude, $J=500, 1000$, and $1500$.

We observe that the distribution of the Gaussian-smoothed FTLE shown in Fig.\ref{fig03}(a) differs slightly from the FTLE distribution in Fig.~\ref{fig02}(a). This difference is not solely due to the coarse graining introduced by the Gaussian smoothing, but also to the way the distributions are sampled. The FTLE distribution in Fig.~2(a) is computed with respect to the natural (Sinai–Bowen–Ruelle)\cite{Yakov_G_Sinai_1972,Ruelle1976AMA,Bowen1975} measure induced by the dynamics, whereas the GFTLE distribution is obtained by sampling uniformly over phase space. This uniform sampling is consistently used for the computation of all GFTLE and PR distributions, ensuring a fair comparison between the classical and quantum indicators.

\subsection{Distributional correlation between PR and GFTLE}

Beyond pointwise correlations, we compare the full probability distributions of PR and coarse-grained FTLE values. The Jensen–Shannon distance provides a symmetric measure of similarity between distributions.

Let $F=\{f_k\}$ and $G=\{g_k\}$ denote two normalized histograms defined over a common support. We introduce the mixed distribution $M=(F+G)/2$, with components $m_k=(f_k +g_k)/2$. The Jensen Shannon divergence  between $F$ and $G$ is defined as

\begin{equation}
\mathrm{JSD}(F|G)
=\frac{1}{2}D_{\mathrm{KL}}(F|M)
+\frac{1}{2}D_{\mathrm{KL}}(G|M),
\end{equation}
where
\[
D_{\mathrm{KL}}(F|M)=\sum_k f_k \ln\!\left(\frac{f_k}{m_k}\right)
\]
is the Kullback--Leibler divergence. The corresponding Jensen--Shannon distance is
\begin{equation}
D_{\mathrm{JS}}(F,G)=\sqrt{\mathrm{JSD}(F|g)},
\end{equation}
which satisfies $0\leq D_{\mathrm{JS}}(F,G)\leq \sqrt{\ln 2}$.

This metric provides a global statistical comparison between the quantum and classical indicators, revealing how closely the hierarchical structures encoded in their distributions align. 

In agreement with the Pearson correlation analysis, we find that the smallest Jensen Shannon distances between classical and quantum indicators 
are located at intermediate time windows, approximately in the range $\tau\simeq 50$--$120$. Outside this range, the Jensen Shannon distance increases, indicating that the PR distributions are either unable to resolve the relevant hierarchical structures at short times or become overly smoothed at long times. ~\cite{quantumrecurrences2024,Pseudoclassical2022}).

\begin{table}[h!]
\centering
\begin{tabular}{||c | c | c||} 
 \hline
 J & $t_{JS}$ & $t_{\rho_{XY}}$ \\ [0.5ex] 
 \hline\hline
 500 \quad & \quad (50:92) \quad & \quad (68:100) \\
 1000 \quad & \quad (56:102) \quad & \quad (90:110) \\
 1500 \quad & \quad (66:94) \quad & \quad (90:120) \\ [1ex] 
 \hline
\end{tabular}
\caption{Time windows $t_{\rho_{XY}}$, around the maximum of the Pearson correlation, and $t_{JS}$, around the minimum of the Jensen Shannon distance, for three values of $J$.}
\label{tab01}
\end{table}

Table \ref{tab01} shows the time windows at which the Jensen--Shannon distance reaches its minimum and the Pearson correlation coefficient reaches its maximum for the three spin magnitudes considered: $J = 500, 1000$, and $1500$.
The existence of an optimal time window where both measures coincide reflects a competition between barrier resolution and ergodic averaging.
\section{Summary and conclusions}

We have shown that hierarchical stickiness in mixed phase space is quantitatively encoded in quantum localization. The participation ratio of coherent states does not merely distinguish regular from chaotic regions; it resolves the layered structure within the chaotic sea that arises from partial transport barriers.

By introducing a Gaussian coarse graining of the finite-time Lyapunov exponent, matched to the intrinsic semiclassical resolution of coherent states, we established a direct quantum–classical correspondence. The agreement is maximized within an intermediate evolution-time window, reflecting a competition between two mechanisms: short times fail to resolve hierarchical trapping, whereas long times average over it. This optimal window identifies the temporal scale at which classical transport structures are most faithfully imprinted onto quantum eigenstates.

Beyond static phase-space indicators, the dynamical evolution of coherent states provides a complementary interpretation of the PR hierarchy: states initialized in regions with smaller FTLE and PR values remain comparatively localized over the evolution times considered, whereas states in more chaotic regions display stronger spreading across phase space. This highlights the importance of considering the hierarchy of stickiness when analyzing localization and transport dynamics in mixed phase-space systems.

The dependence of the correlation on spin size confirms the semiclassical origin of this effect: as $\hbar_{\mathrm{eff}}\rightarrow 0$, quantum localization increasingly mirrors the classical hierarchy. The correspondence is therefore consistent with the phase-space organization of mixed systems. More generally, our results expand the scope of the participation ratio as a diagnostic tool in mixed systems. Beyond its established role in identifying fixed points, quantum scars, and symmetry-related structures, we show that the PR can also resolve hierarchical sticky regions within the chaotic sea whenever partial transport barriers become dynamically relevant. In this sense, the participation ratio provides information not only about the global regular–chaotic transition, but also about the internal organization of transport in mixed phase spaces.

The persistence of classical quantum correspondence in sticky regions suggests that, in mixed phase-space systems, effective Ehrenfest like times may depend on the local finite-time instability rather than solely on the asymptotic Lyapunov exponent. This possibility will be explored in future work.

More broadly, the present framework can be extended to other Floquet systems and to higher-dimensional or many-body settings where hierarchical and/or anomalous transport structures play a central role.

\begin{acknowledgments}
We acknowledge the support of the Computation Center - ICN to develop many of the results presented in this work, in particular to Enrique Palacios, Luciano Díaz, and Eduardo Murrieta. 
This work received partial financial support form DGAPA- UNAM projects IN109523 and IN101526.

\section*{AUTHOR DECLARATIONS}
\subsection*{Conflict of Interest}
The authors have no conflicts to disclose.

\subsection*{Author Contributions}

\textbf{Ariel A. Galindo Duque}: Conceptualization (equal); Investigation (equal); Methodology (equal); Software (equal); Writing – review (equal). \textbf{Miguel A. Prado Reynoso}: Conceptualization (equal); Methodology (equal); Supervision (supporting); Writing – original draft (lead). \textbf{Miguel Gonzalez}: Conceptualization (supporting); Methodology (equal); Software (equal); Writing – review (equal). \textbf{Jorge G. Hirsch}: Supervision (lead); Conceptualization (equal); Formal analysis(lead); Methodology (equal); Writing – review and editing (lead).

\end{acknowledgments}

\appendix
\section{Classical and quantum kicked top} 
\label{ap1}
The quantum kicked top is defined by the Floquet operator (\ref{FO}).
To probe the quantum dynamics locally in phase space, we employ spin-coherent states
\begin{equation}
|\theta, \phi\rangle = 
e^{i \theta \left( \hat{J}_y \sin \phi - \hat{J}_x \cos \phi \right)} |J, J\rangle,
\end{equation}
which are minimal-uncertainty states over the surface of the Bloch sphere centered at the point $(\theta,\phi)$. Their resolution scales as $\Delta \theta \approx \Delta \phi \approx J^{1/2}$.

Taking the expectation value of the angular momentum operators with respect to spin-coherent states allows to introduce the classical variables 
$$
\lim_{\!J\rightarrow\infty\!} \langle \theta, \phi | \hat{\mathbf{J}} |\theta, \phi\rangle = \vec{J}=\left(J_x, J_y, J_z\right).
$$

The classical Hamiltonian is obtained from the quantum Hamiltonian (\ref{Hq}) by taking the expectation value with respect to spin-coherent~\cite{PhysRevB.104.104409} states and then taking the limit $J\rightarrow\infty$, as follows:
\begin{equation}
\begin{split}
H \!\!=\!\! &\lim_{\!J\rightarrow\infty\!}\left\{\frac{\langle \hat{H}(t)\rangle_{\theta,\phi}}{J}\right\} 
\!=\! \alpha J_z \!+ \frac{k}{2} J_x^2 \sum_{n=-\infty}^\infty \!\!\!\!\delta\!\left(t - n\right)
\end{split}
\label{Hcl}
\end{equation}

We parametrize the classical spherical phase space $\mathbf{S}_2$ by its stereographic coordinates
\begin{equation}
Q=\frac{\sqrt{2}J_x}{\sqrt{1-J_z}}, \qquad
P=\frac{\sqrt{2}J_y}{\sqrt{1-J_z}},
\label{eq:canonicalpair}
\end{equation}
where the projection sends the sphere onto the disk $Q^2 + P^2 <4$. Using this parametrization, the associated spin coherent state takes the form
\begin{equation}
\label{spin_c}
\big|z(Q,P)\rangle=\frac{e^{z\hat{J}_+}}{\left( 1 + |z|\right)^J}|J,-J\rangle
\end{equation}
where $z=(Q+iP)/\sqrt{4-(Q^2+p^2)}$ and $\hat{J}_+=\hat{J}_x+i\hat{J}_y$ is the creation operator. 

\section{Participation Ratio Across the Regular–Chaotic Transition}
\label{ap2}

\begin{figure*}[th]
    \centering
    \includegraphics[width=0.9\linewidth]{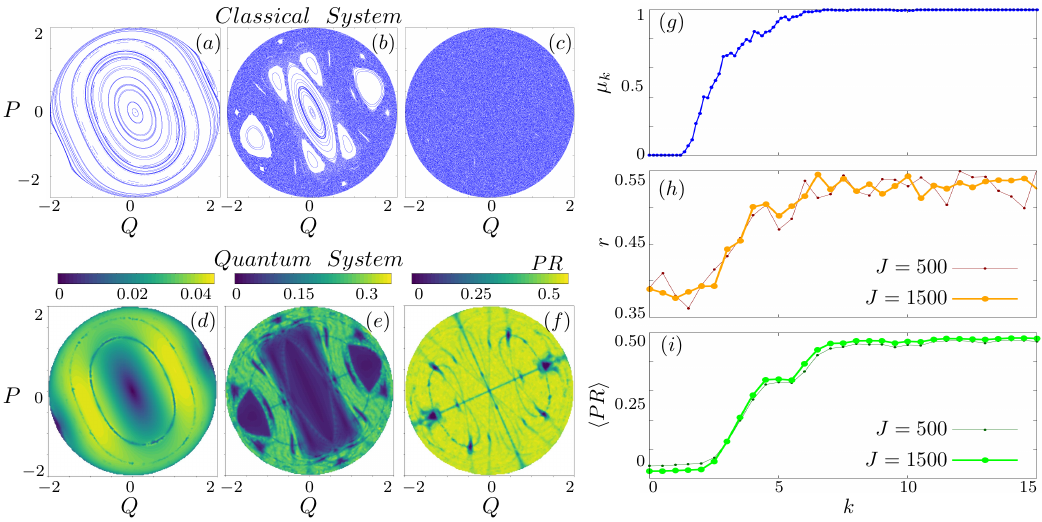}
    \caption{
(a–c) Classical phase space for increasing kick strength $k=1, 4$, and $10$, showing the progressive destruction of invariant tori and the growth of the chaotic sea. (d–f) Quantum counterparts obtained from the participation ratio of coherent states evaluated over phase space for the same parameter values. (g) Fraction of chaotic phase-space area relative to the regular region as a function of the kick strength. (h) Level-spacing parameter characterizing the spectral transition from regular to chaotic statistics. (i) Phase-space–averaged participation ratio as a function of the kick strength.}
    \label{fig04}
\end{figure*}

In this section we  analyze the regular-chaotic transition by jointly examining the evolution of classical phase space structures and their quantum counterparts as the kicking strength $k$ is varied, while keeping the value $\alpha=0.84$ fixed. This transition is consistently captured by three complementary indicators:
the classical chaotic area fraction $\mu_k$, the quantum quasienergy level-spacing statistic $r$, and the phase space averaged participation ratio $\langle PR\rangle$. 
The latter captures the same transition through the localization properties of coherent states, without relying on spectral statistics. These indicators are shown in Figs.~\ref{fig04}(g), \ref{fig04}(h), and \ref{fig04}(i), respectively.

For each value of $k$, we generate $1000$ random initial conditions and compute the Lyapunov spectrum over an evolution time of $t=10^6$. For the quantum calculations, wepresent results for two angular momentum magnitudes: $J=500$ and $J=1500$.
It can be seen that both values provide 
a well-defined semiclassical regime.

At weak kicking strengths, $k\lesssim 2.5$, the dynamics remain predominantly regular. Phase space is largely filled with invariant tori and stable periodic orbits, as illustrated in Fig.~\ref{fig04}(a), and the chaotic fraction $\mu_k$ is close to zero. In this regime, the quasienergy spectrum exhibits Poisson-like statistics, reflected by small values of the level-spacing parameter $r$, while the phase-space--averaged participation ratio remains low. Consistently, the corresponding PR landscapes display strongly localized coherent states, indicating the persistence of classical regularity in the quantum dynamics (see Fig.~\ref{fig04}(d). Notice that the largest value of the PR is very small: $0.04$).

As $k$ increases into the intermediate range $2.5\lesssim k\lesssim 4$, the system enters a mixed dynamical regime in which regular islands coexist with chaotic layers, as shown in Fig.~\ref{fig04}(b). The breakup of separatrices leads to heteroclinic and homoclinic tangles, generating a hierarchical organization of phase space and long-lived trapping of chaotic trajectories in sticky regions~\cite{WangRobnik2021,Wang2021phase}. In this regime, the chaotic area fraction $\mu_k$ grows steadily, while the level-spacing statistic $r$ departs from Poisson behavior but has not yet reached its fully chaotic limit. At the same time, the average participation ratio increases, and the PR landscapes exhibit intermediate values, signaling partial delocalization of coherent states. As discussed in the following section, this behavior constitutes a clear quantum manifestation of classical stickiness (see Fig.~\ref{fig04}(e)).

For sufficiently strong kicking, $k>4$, the dynamics become globally chaotic, with phase space almost entirely covered by chaotic trajectories, as illustrated in Fig.~\ref{fig04}(c). Only small stability islands survive, and their phase-space measure rapidly approaches zero, leading to saturation of the chaotic fraction $\mu_k$. In this regime, the level-spacing statistic $r$ approaches the predictions of random-matrix theory, and the phase-space-averaged participation ratio attains uniformly high values. Accordingly, coherent states are almost completely delocalized over the Floquet eigenbasis, and the PR is nearly uniform throughout phase space (see Fig.~\ref{fig04}(f)).

Notably, this near-uniformity is punctuated by a set of thin, curve-like structures along which the PR attains values that are slightly lower than the typical values observed in this regime. Although the PR remains high overall, these filaments stand out against the otherwise homogeneous background and signal localized deviations from complete delocalization. While their detailed origin and dynamical significance are not addressed here, their visibility suggests the persistence of correlations and symmetries in phase space. A dedicated analysis of these low-PR curves will be presented in future work.

\section{Lyapunov exponents}
\label{ap3}

For an invertible, measure-preserving map $f:M\to M$, the asymptotic Lyapunov exponent associated with an initial condition $\mathbf{x}$ and tangent vector $\mathbf{u}$ is defined as
\begin{equation}
\lambda=\lim_{t\to\infty}\frac{1}{t}
\log\bigl\|D_{\mathbf{x}}f^{t}(\mathbf{u})\bigr\|,
\label{LE}
\end{equation}
where $\lambda>0$ signals chaotic behavior.

The FTLE measures the local divergence rate of nearby trajectories over a finite time window, it is particularly sensitive to sticky dynamics: trajectories that linger near regular islands exhibit temporarily reduced instability, resulting in anomalously small FTLE values. Spatial variations of the FTLE therefore provide a natural diagnostic of the hierarchical structures that surround regular islands in mixed phase space.

The FTLE follows directly from the invariance of the unstable tangent space underlying the asymptotic Lyapunov exponent introduced in Eq.~(\ref{LE}). For a unit tangent vector $v_{\mathbf{x}}^{u}$ aligned with the corresponding unstable direction, the FTLE associated with the trajectory $\{ f^{\tau}(\mathbf{x}) \}$ of time $\tau$, is defined as
\begin{equation}
\lambda_{\tau,\mathbf{x}}
    = \frac{1}{\tau}\log\|D_{\mathbf{x}}f^{\tau}(v_{\mathbf{x}}^{\, u})\|.
\end{equation}
In the limit $\tau \to \infty$, the $FTLE$ converges to the corresponding asymptotic Lyapunov exponent. In strictly regular regions, however, no unstable direction exists; consequently, the FTLE is not defined there.

\section{Gaussian-smoothed finite time Lyapunov exponent (GFTLE).}
\label{ap4}

Let $\lambda_{\tau,z}$ denote the FTLE evaluated at the point $z=(\theta,\varphi)\in\mathbb{S}^2$. For a reference point $x=(\theta_0,\varphi_0)$ at which we wish to compute a local classical instability measure, we define the Gaussian smoothed FTLE (GFTLE) as
\begin{equation}
\Lambda_{\tau,x}
=
\frac{
\displaystyle
\int_{\mathbb{S}^2} 
\lambda_{\tau,z} \,
\exp\!\left[
-\frac{\gamma(z,x)^2}{2\sigma^2}
\right]
\, d\Omega
}{
\displaystyle
\int_{\mathbb{S}^2} 
\exp\!\left[
-\frac{\gamma(z,x)^2}{2\sigma^2}
\right]
\, d\Omega
},
\qquad
\sigma^2=\frac{1}{J}.
\end{equation}
Here $\gamma(z,x)=\arccos(\cos\theta\cos\theta_0+\sin\theta\sin\theta_0\cos(\varphi-\varphi_0))$ is the geodesic distance on $\mathbb{S}^2$, and $d\Omega=\sin\theta\,d\theta\,d\varphi$ is the standard surface element. The choice $\sigma^2=1/J$ is consistent with the angular spread of spin coherent states and ensures that the classical coarse graining matches the intrinsic quantum resolution set by $\hbar_{\mathrm{eff}}$.

Importantly, this smoothing is performed directly in the angular-momentum phase-space variables, i.e., on the spherical manifold $\mathbb{S}^2$ where the classical kicked-top dynamics is defined.
\bibliography{refs.bib}
\bibliographystyle{apsrev4-2}

\end{document}